\documentclass[review]{elsarticle}

\usepackage{lineno,hyperref}
\usepackage{amsmath}
\usepackage{amssymb}
\usepackage{yhmath}
\usepackage{booktabs, multirow} 
\usepackage{graphicx,psfrag,epsf}
\usepackage{float}
\usepackage{caption}
\usepackage{subcaption} 
\graphicspath{ {./figures/} }
\usepackage{xcolor} 
\usepackage{hyperref}
\hypersetup{colorlinks,
citecolor=blue
}
\usepackage{geometry}
\makeatletter
\def\ps@pprintTitle{%
  \let\@oddhead\@empty
  \let\@evenhead\@empty
  \let\@oddfoot\@empty
  \let\@evenfoot\@oddfoot
}
\makeatother

\modulolinenumbers[5]

\begin{document}
\include{macros} 

\begin{frontmatter}

\title{Functional Connectivity:\\ Continuous-Time Latent Factor Models for Neural Spike Trains}

\author[mainaddress]{Meixi Chen}
\author[mainaddress]{Martin Lysy}
\author[secondaddress]{David Moorman}
\author[mainaddress]{Reza Ramezan\corref{mycorrespondingauthor}}
\ead{rramezan@uwaterloo.ca}
\cortext[mycorrespondingauthor]{Corresponding author}

\address[mainaddress]{Department of Statistics and Actuarial Science, University of Waterloo, Canada}
\address[secondaddress]{Department of Psychological and Brain Sciences, University of Massachusetts Amherst, USA}

\begin{abstract}
Modelling the dynamics of interactions in a neuronal ensemble is an important problem in functional connectivity research. One popular framework is latent factor models (LFMs), which have achieved notable success in decoding neuronal population dynamics. However, most LFMs are specified in discrete time, where the choice of bin size significantly impacts inference results. In this work, we present what is, to the best of our knowledge, the first continuous-time multivariate spike train LFM for studying neuronal interactions and functional connectivity.  We present an efficient parameter inference algorithm for our biologically justifiable model which (1) scales linearly in the number of simultaneously recorded neurons and (2) bypasses time binning and related issues. Simulation studies show that parameter estimation using the proposed model is highly accurate. Applying our LFM to experimental data from a classical conditioning study on the prefrontal cortex in rats, we found that coordinated neuronal activities are affected by (1) the onset of the cue for reward delivery, and (2) the sub-region within the frontal cortex (OFC/mPFC). These findings shed new light on our understanding of cue and outcome value encoding.
\medskip
\end{abstract}


\end{frontmatter}

\section{Introduction}

An important question in neuroscience research is understanding the functional connectivity between neurons in different parts of the brain. Spike trains based on simultaneously recorded neurons provide information about population coding and neuronal interaction.
Both model-free and model-based spike-train analysis tools have been developed to answer this question.
While model-free methods \citep{perkel-etal67, ventura-etal05, fujisawa08, humphries11, lopes-dos-santos-etal13} are typically more efficient and convenient to implement, they often fail to uncover more complex underlying neuronal relationships beyond correlation at the level of observed data. In contrast, latent factor models (LFMs) are able to discover patterns which model-free algorithms cannot, thanks to their ability to specify different structures in the latent layers. 
The remarkable success of spike-train LFMs in predictive tasks \citep{yu-etal08, gao-etal16, wu-etal17, pandarinath-etal18} and neuronal clustering \citep{buesing-etal14, wei-etal2022} motivates our work in this paper to study the functional connectivity between neurons. As all the aforementioned LFMs are specified in discrete time, they are commonly applied upon binning the experimental time to obtain spike counts, which are then modelled using, e.g., a Poisson likelihood. However, the bin size is often chosen arbitrarily-- despite this having a significant impact on parameter estimation \citep{nelson02, kass-ventura06, ramezan-etal14}. Although there are Poisson process models set in continuous time without binning \citep[e.g.,][]{duncker-sahani-18,williams-etal20}, these models cannot be readily applied to the functional connectivity analysis of multiple neurons.\\

In this work we present a continuous-time multivariate point process LFM to study neuronal interactions based on simultaneously recorded spike trains in a neural  population. To the best of our knowledge, this is the first continuous-time LFM proposed for the analysis of neural spike trains. In our model, the activities of a neuronal population are described by correlated Wiener processes with resetting.  Each of these processes is viewed as a proxy of the evolving membrane potential of a neuron which resets after reaching a threshold.  Crucially, we assume that the high-dimensional multivariate latent process can be summarized by a small number of dynamic factors.  Not only does this factor analysis framework provide an interpretable low-dimensional representation of neural activities, it also serves as a means of dimension reduction for studying large neuronal populations.  Our model generalizes the limiting case of the multivariate Skellam point process with resetting of \cite{ramezan-etal22}.  However, by passing to the Brownian limit, we are able to develop an efficient algorithm for parameter inference, which both circumvents the choice of bin size (it can be made so small as to approximate the continuous time process arbitrarily well) and scales linearly in the number of neurons in the analysis.  The applicability of the proposed model and inference procedure is demonstrated in simulated and experimental data analyses.

\begin{figure*}[ht]
\vspace*{1em}
\centering
\includegraphics[width=\linewidth]{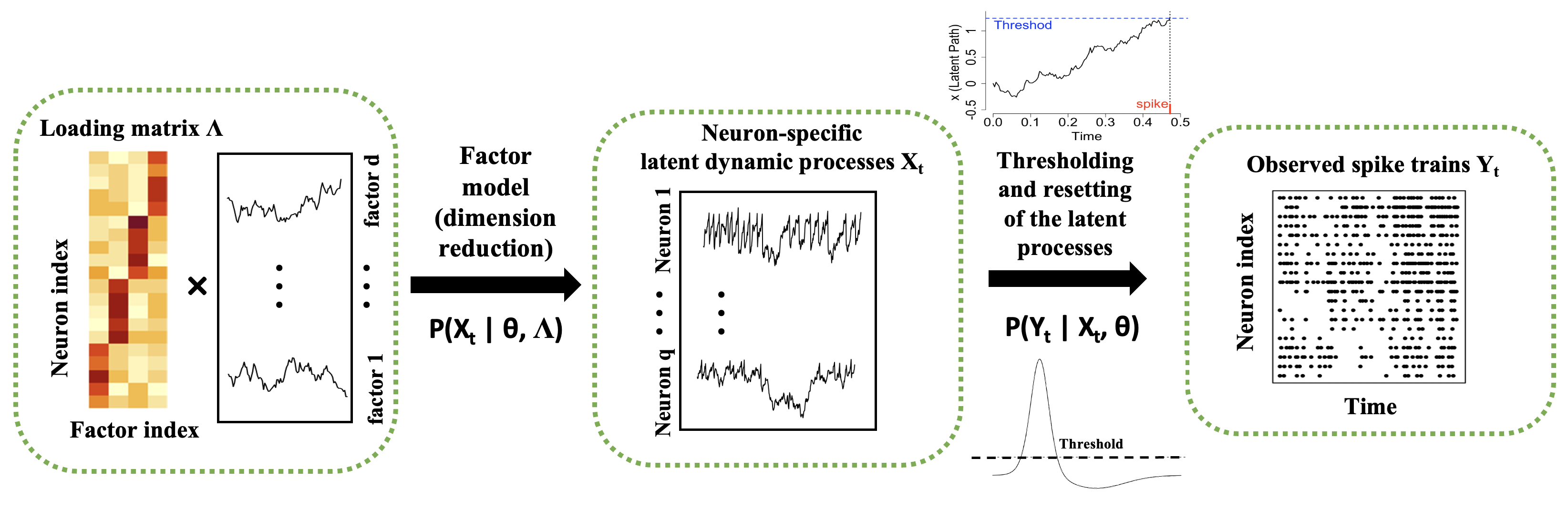}
\vspace*{-3em}
\caption{Illustration of the proposed model.} \label{fig:model}
\end{figure*}

\section{Model}
A graphical summary of the proposed model is presented in Figure~\ref{fig:model}. Let $\mathbf{Y}_t= (y_{1,t},\ldots,y_{q,t})\in\{0,1\}^q$ denote the observed binary spike trains of a population of $q$ neurons at time $t$, and $\mathbf{X}_t = (x_{1,t},\ldots,x_{q,t}) \in\mathbb{R}^q$ be the unobserved latent dynamic processes of the neuronal population. We assume that the neuronal activities are governed by $\mathbf{X}_t$, which are modelled as correlated Wiener processes with resetting. Each process $i$ is reset to its initial value whenever it crosses some neuron-specific threshold, and then a spike occurs, i.e., $y_{i,t}=1$ is observed. We let $\boldsymbol{\theta}$ denote the threshold parameters and the Wiener process drift vector. The threshold-crossing and resetting processes mimic the spike-generating mechanism and the refractoriness of the neuron, respectively.  We further assume that the $q$ dynamic processes can be represented as $\mathbf{X}_t = \boldsymbol{\Lambda} \mathbf{F}_t + \boldsymbol{\epsilon}_t$, where $\mathbf{F}_t = (f_{1,t}, \ldots, f_{d,t})$ are $d \ (d\ll q)$ dynamic factors, and $\boldsymbol{\epsilon}_t = (\epsilon_{1,t}, \ldots, \epsilon_{q,t})$ are neuron-specific idiosyncrasies.  The loading matrix $\boldsymbol{\Lambda}$ can be interpreted within the traditional factor analysis framework-- that is, it identifies a small number of factors driving the neuronal dynamic processes.  It is also used to model the correlation matrix between latent neuron processes, i.e., $\operatorname{cor}(\mathbf{X}_t) = \boldsymbol{\Sigma} = \boldsymbol{\Lambda} \boldsymbol{\Lambda}' + \boldsymbol{\Psi}$, where $\boldsymbol{\Psi}$ is a diagonal matrix with elements determined by $\boldsymbol{\Lambda}$ and the fact that $\boldsymbol{\Sigma}$ has unit diagonal.

\section{Inference}
Model inference is carried out in two steps. First, the drift of the Wiener process and the threshold parameters, summarized by $\boldsymbol{\theta}$, can be estimated analytically since the first passage time of a Wiener process follows an Inverse Gaussian distribution \citep{brown05}. However, closed-form estimation of $\boldsymbol{\Lambda}$ is not available. One common solution is to employ MCMC sampling on $p(\mathbf{X}_{1:T}, \boldsymbol{\Lambda} \mid \mathbf{Y}_{1:T})$, where time has been discretized to an arbitrarily fine grid.  However, this can be prohibitively slow in high dimensions. Instead, the latent processes $\mathbf{X}_{1:T}$ are integrated out via the Laplace approximation \citep{tierney-kadane86, skaug.fournier06, Koyama2010}, i.e., $p_{\mathrm{LA}}(\mathbf{Y}_{1:T}\mid \boldsymbol{\Lambda}) \approx \int p(\mathbf{Y}_{1:T},\mathbf{X}_{1:T}\mid\boldsymbol{\Lambda})\ d\mathbf{X}_{1:T}$, so that an estimate of $\boldsymbol{\Lambda}$ can be obtained by maximizing the Laplace-approximated marginal likelihood $p_{\mathrm{LA}}(\mathbf{Y}_{1:T}\mid \boldsymbol{\Lambda})$.  An efficient gradient-based nested optimization algorithm for this is implemented in our \textsf{R}/\textsf{C++} library \texttt{fastr} \citep{chen-etal23}.  The algorithm involves repeatedly solving linear systems of the form $\boldsymbol{\Sigma}\mathbf{x} = \mathbf{b}$, which due to the factor structure $\boldsymbol{\Sigma} = \boldsymbol{\Lambda} \boldsymbol{\Lambda}' + \boldsymbol{\Psi}$, scales linearly in the number of neurons $q$ for fixed number of factors $d$.

\section{Data Analysis}
\begin{figure}[ht]
\centering
\includegraphics[width=.8\linewidth]{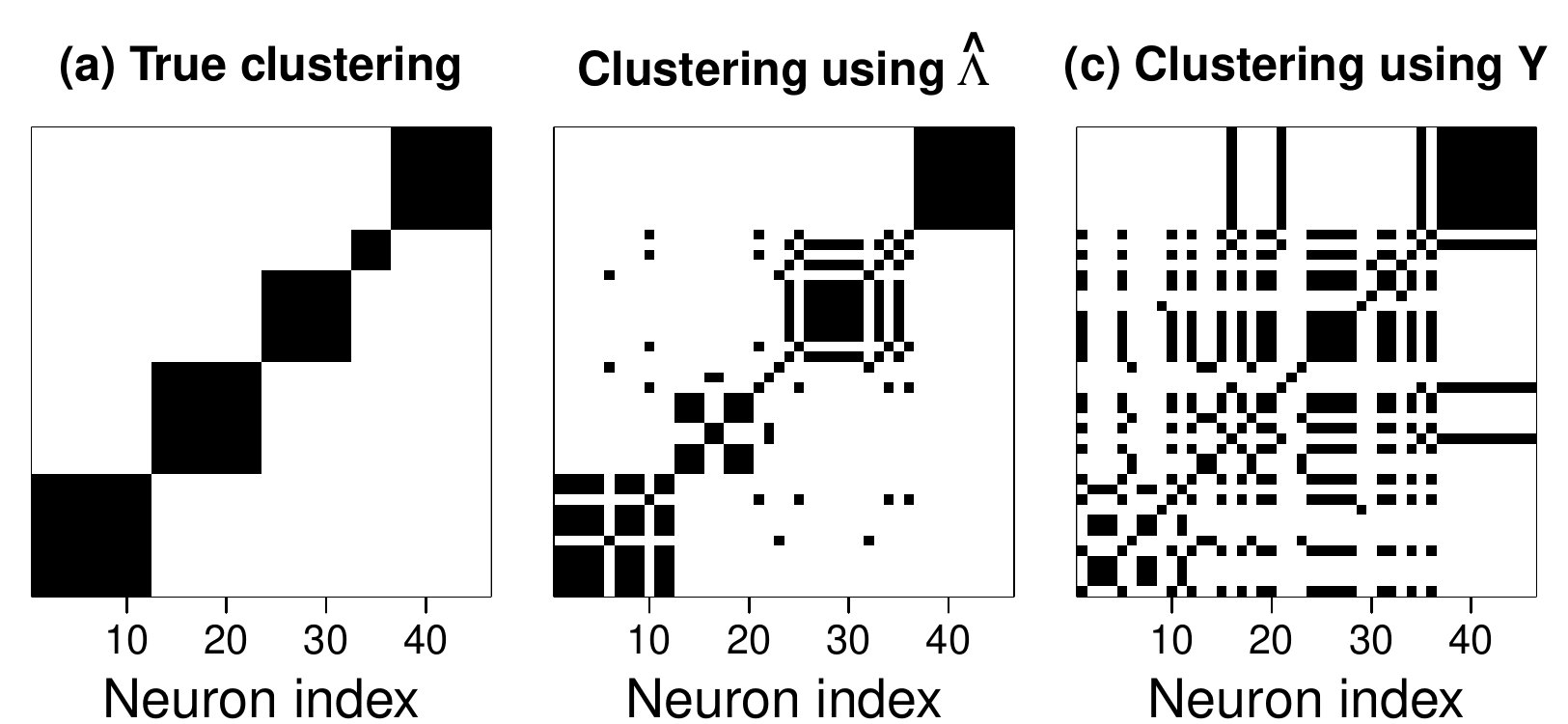}
\caption{Clustering performance. The $(i,j)$-th entry of the matrix is black if neuron $i$ and $j$ are assigned to the same cluster, and white otherwise.} \label{fig:sim}
\vspace*{-0.7em}
\end{figure}

\begin{figure}[ht]
\centering
\includegraphics[scale=0.5]{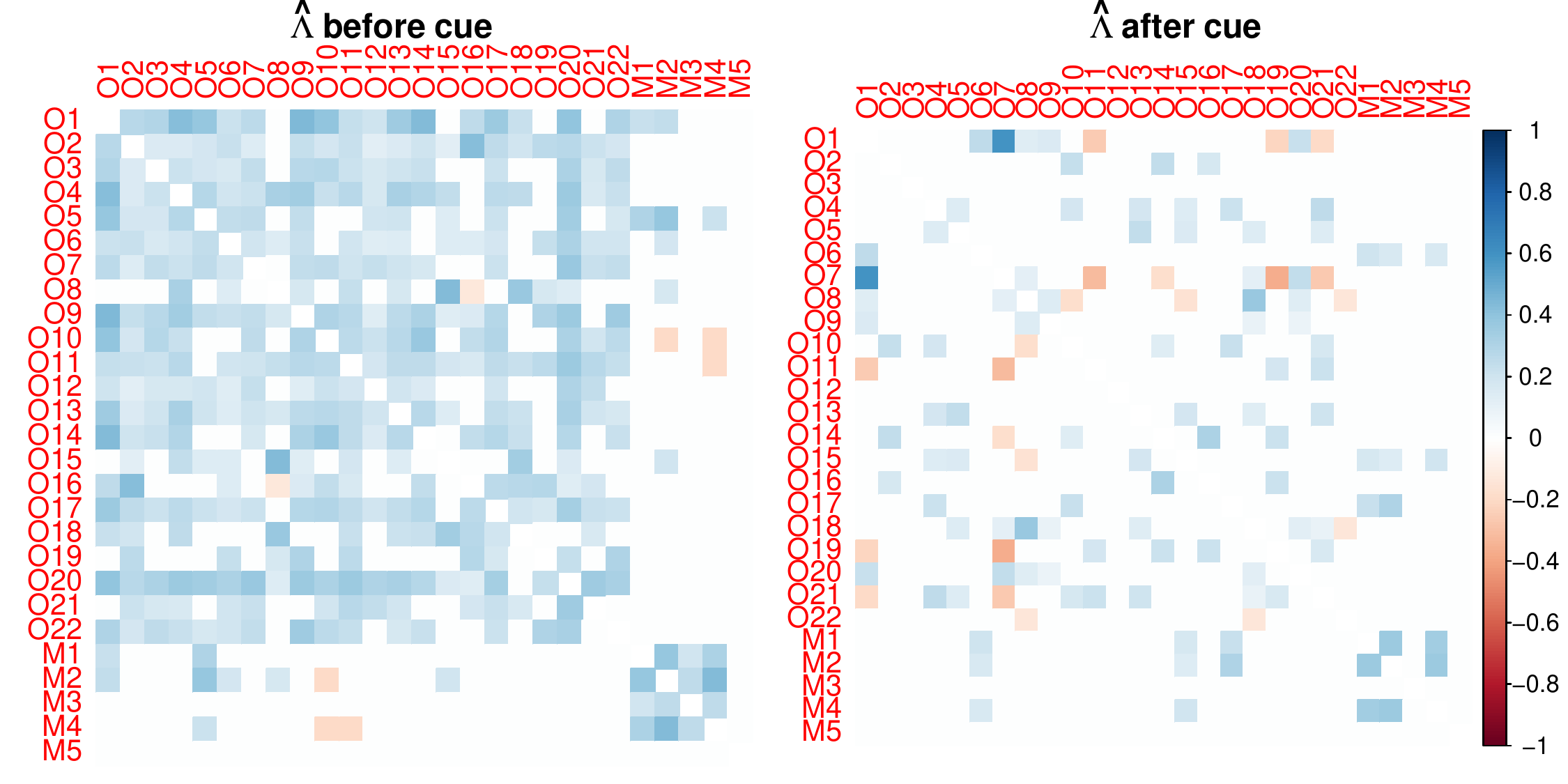}
\caption{Estimated correlation between neurons. "O" stands for OFC neurons and "M" stands for mPFC neurons. The numbers are the neuron indices. For visual presentation, non-significant values and the diagonal elements are set to zero.} \label{fig:cor}
\vspace*{-1em}
\end{figure}

To assess the performance of the proposed model, we first applied it to a biologically plausible simulated dataset generated in NEURON \citep{hines-carnevale97}. We found that the $K$-means clustering results based on $\hat{\boldsymbol{\Lambda}}$ are far more accurate than those based on (convolved) multiple neuron spike trains $\mathbf{Y}_{1:T}$ (see Figure~\ref{fig:sim}(a)-(c)). Therefore, the estimate $\hat{\boldsymbol{\Lambda}}$ uncovers the underlying associations between neurons beyond what is observed at the data level. Next, we applied our model on neuronal ensembles recorded from a rat's medial prefrontal cortex (mPFC) and orbitofrontal cortex (OFC) during a classical conditioning experiment. The rat was trained to recognize a cue predictive of an appetitive outcome (sucrose). We modelled neuronal population dynamics in relation to encoding appetitive outcomes and functional connectivity, and found that coordinated activities are modulated by brain areas mPFC and OFC. The estimated correlation matrices $\hat{\boldsymbol{\Sigma}}$ in Figure~\ref{fig:cor} also show a decrease in overall neuronal interactions after the onset of the cue. 

\section{Discussion}
We have proposed a novel continuous-time multivariate point process latent factor model for simultaneously recorded spike trains. Downstream analyses using the proposed model can reveal neuronal clustering and estimate correlations between neurons. Computational challenges in model inference are addressed by carefully designing and implementing an efficient model-fitting procedure based on the Laplace approximation. We have confirmed, via simulation studies, that our algorithm achieves satisfactory accuracy and precision in parameter estimation. Finally, compared to black-box deep learning models for neural spike trains, our proposed model is able to provide more reliable statistical inference results with uncertainty quantification, which is vital for making scientifically sound conclusions. One immediate future direction is to apply our model to appetitive and aversive classical conditioning outcomes (available within the same experimental data) to investigate reward-value coding, and to identify value signals for the most relevant contextual features.

\section*{Acknowledgments}
This work was supported by the Natural Sciences and Engineering Research Council of Canada, grant numbers RGPIN-2018-04376 (Ramezan), DGECR-2018-00349 (Ramezan) and RGPIN-2020-04364 (Lysy).\\

\section*{References}

\bibliographystyle{plain}

\bibliography{ChenEtAl2023Bib}

\begin{thebibliography}{10}

\bibitem{brown05}
E.~N Brown.
\newblock Theory of point processes for neural systems.
\newblock In {\em Methods and Models in Neurophysics}, chapter~14, pages
  691--727. Elsevier, 2005.

\bibitem{buesing-etal14}
L.~Buesing, T.~A. Machado, J.~P. Cunningham, and L.~Paninski.
\newblock Clustered factor analysis of multineuronal spike data.
\newblock In {\em Advances in Neural Information Processing Systems},
  volume~27, 2014.

\bibitem{chen-etal23}
Meixi Chen, Reza Ramezan, and Martin Lysy.
\newblock fastr: {\textbf{F}actor \textbf{A}nalysis of \textbf{S}pike
  \textbf{T}rains in \textbf{R}}.
\newblock \url{https://github.com/meixichen/fastr}, 2023.

\bibitem{duncker-sahani-18}
Lea Duncker and Maneesh Sahani.
\newblock Temporal alignment and latent {Gaussian} process factor inference in
  population spike trains.
\newblock In {\em Advances in Neural Information Processing Systems}, 2018.

\bibitem{fujisawa08}
S.~Fujisawa, A.~Amarasingham, M.~T. Harrison, and G.~Buzsáki.
\newblock Behavior-dependent short-term assembly dynamics in the medial
  prefrontal cortex.
\newblock {\em Nature Neuroscience}, 11:823–833, 2008.

\bibitem{gao-etal16}
Yuanjun Gao, Evan Archer, Liam Paninski, and John~P. Cunningham.
\newblock Linear dynamical neural population models through nonlinear
  embeddings.
\newblock In {\em 30th Conference on Neural Information Processing Systems},
  2016.

\bibitem{hines-carnevale97}
M.~L. Hines and N.~T. Carnevale.
\newblock The neuron simulation environment.
\newblock {\em Neural Computation}, 9:1179--1209, 1997.

\bibitem{humphries11}
M.~D. Humphries.
\newblock Spike-train communities: Finding groups of similar spike trains.
\newblock {\em Journal of Neuroscience}, 31(6):2321--2336, 2011.

\bibitem{kass-ventura06}
Robert~E. Kass and Valérie Ventura.
\newblock Spike count correlation increases with length of time interval in the
  presence of trial-to-trial variation.
\newblock {\em Neural Computation}, 18(11):2583--2591, 2006.

\bibitem{Koyama2010}
S.~Koyama, U.~T. Eden, E.N. Brown, and R.E. Kass.
\newblock Bayesian decoding of neural spike trains.
\newblock {\em Annals of the Institute of Statistical Mathematics}, 62:37--59,
  2010.

\bibitem{lopes-dos-santos-etal13}
V.~{Lopes-dos-Santos}, S.~Ribeiro, and A.~B.~L. Tort.
\newblock Detecting cell assemblies in large neuronal populations.
\newblock {\em Journal of Neuroscience Methods}, 220:149--166, 2013.

\bibitem{nelson02}
Mark~E. Nelson.
\newblock Multiscale spike train variability in primary electrosensory
  afferents.
\newblock {\em Journal of Physiology-Paris}, 96(5):507--516, 2002.

\bibitem{pandarinath-etal18}
Chethan Pandarinath, Daniel~J. O’Shea, Jasmine Collins, Rafal J{\'o}zefowicz,
  Sergey~D. Stavisky, Jonathan~C. Kao, Eric~M. Trautmann, Matthew~T. Kaufman,
  Stephen~I. Ryu, Leigh~R. Hochberg, Jaimie~M. Henderson, Krishna~V. Shenoy,
  L.~F. Abbott, and David Sussillo.
\newblock Inferring single-trial neural population dynamics using sequential
  auto-encoders.
\newblock {\em Nature Methods}, 15(10):805--815, 2018.

\bibitem{perkel-etal67}
Donald~H. Perkel, George~L. Gerstein, and George~P. Moore.
\newblock Neuronal spike trains and stochastic point processes. ii.
  simultaneous spike trains.
\newblock {\em Biophysical journal}, 7(5):419--40, 1967.

\bibitem{ramezan-etal22}
Reza Ramezan, Meixi Chen, Martin Lysy, and Paul Marriott.
\newblock A multivariate point process model for simultaneously recorded neural
  spike trains.
\newblock In {\em Conference on Cognitive Computational Neuroscience}, 2022.

\bibitem{ramezan-etal14}
Reza Ramezan, Paul Marriott, and Shoja'eddin Chenouri.
\newblock Multiscale analysis of neural spike trains.
\newblock {\em Statistics in Medicine}, 33:238–256, 2014.

\bibitem{skaug.fournier06}
Hans~J Skaug and David~A Fournier.
\newblock Automatic approximation of the marginal likelihood in non-gaussian
  hierarchical models.
\newblock {\em Computational Statistics \& Data Analysis}, 51(2):699--709,
  2006.

\bibitem{tierney-kadane86}
L.~Tierney and J.~Kadane.
\newblock Accurate approximations for posterior moments and marginal densities.
\newblock {\em Journal of the American Statistical Association}, 81:82--86,
  1986.

\bibitem{ventura-etal05}
Val{\'e}rie Ventura, Can Cai, and Robert~E. Kass.
\newblock Trial-to-trial variability and its effect on time-varying dependency
  between two neurons.
\newblock {\em Journal of neurophysiology}, 94(4):2928--2939, 2005.

\bibitem{wei-etal2022}
Ganchao Wei, Ian~H. Stevenson, and Xiaojing Wang.
\newblock Bayesian clustering of neural activity with a mixture of dynamic
  {Poisson} factor analyzers, 2022.

\bibitem{williams-etal20}
Alex~H. Williams, Anthony Degleris, Yixin Wang, and Scott~W. Linderman.
\newblock Point process models for sequence detection in high-dimensional
  neural spike trains.
\newblock In {\em Advances in Neural Information Processing Systems}, 2020.

\bibitem{wu-etal17}
Anqi Wu, Nicholas~A. Roy, Stephen Keeley, and Jonathan~W Pillow.
\newblock Gaussian process based nonlinear latent structure discovery in
  multivariate spike train data.
\newblock In {\em Advances in Neural Information Processing Systems},
  volume~30, 2017.

\bibitem{yu-etal08}
Byron~M. Yu, John~P. Cunningham, Gopal Santhanam, Stephen~I. Ryu, Krishna~V.
  Shenoy, and Maneesh Sahani.
\newblock Gaussian-process factor analysis for low-dimensional single-trial
  analysis of neural population activity.
\newblock {\em Journal of Neurophysiology}, 102(1):614--635, 2008.

\end{thebibliography}

\end{document}